\documentclass[conference]{IEEEtran}
\usepackage{graphicx}
\usepackage{epstopdf}
\usepackage[latin1]{inputenc}
\usepackage{amsmath,amssymb,amscd,latexsym,dsfont}
\usepackage[english]{babel}
\usepackage{tabularx,cite}
\usepackage{graphicx}
\usepackage{algpseudocode}
\usepackage{algorithm}
\usepackage{algorithmicx}
\usepackage{float}
\usepackage{multicol}
\usepackage{psfrag}
\usepackage{comment,psfrag,subfigure,enumerate}
\usepackage{color}
\usepackage{amsthm}
\usepackage{graphicx}
\usepackage{graphicx}
\usepackage{epstopdf}
\usepackage[latin1]{inputenc}
\usepackage{amsmath,amssymb,amscd,latexsym,dsfont}
\usepackage[english]{babel}
\usepackage{tabularx,cite}
\usepackage{graphicx}
\usepackage{algpseudocode}
\usepackage{algorithm}
\usepackage{algorithmicx}
\usepackage{float}
\usepackage{multicol}
\usepackage{mathtools}
\usepackage{psfrag}
\usepackage{comment,psfrag,subfigure,enumerate}
\usepackage{color}
\usepackage{amsthm}
\ifCLASSINFOpdf
\else
\fi
\begin{document}
\IEEEoverridecommandlockouts
\IEEEpubid{\makebox[\columnwidth]{978-1-4799-5863-4/14/\$31.00 \copyright 2014 IEEE \hfill} \hspace{\columnsep}\makebox[\columnwidth]{ }}
\title{Performance analysis of RF-FSO multi-hop networks}

\author{
    \IEEEauthorblockN{Behrooz Makki\IEEEauthorrefmark{1}, Tommy Svensson\IEEEauthorrefmark{1}, Maite Brandt-Pearce\IEEEauthorrefmark{2} and Mohamed-Slim Alouini\IEEEauthorrefmark{3}}
    \IEEEauthorblockA{\IEEEauthorrefmark{1}Chalmers University of Technology, Gothenburg, Sweden, \{behrooz.makki, tommy.svensson\}@chalmers.se}
\IEEEauthorblockA{\IEEEauthorrefmark{2}University of Virginia, Charlottesville, VA , USA, mb-p@virginia.edu}
    \IEEEauthorblockA{\IEEEauthorrefmark{3} King Abdullah University of Science and Technology (KAUST), Thuwal, Saudi Arabia, slim.alouini@kaust.edu.sa}
}

%


\maketitle
\vspace{-0mm}
\begin{abstract}
We study the performance of multi-hop networks composed of millimeter wave (MMW)-based radio frequency (RF) and free-space optical (FSO) links. The results are obtained in the cases with and without hybrid automatic repeat request (HARQ). Taking the MMW characteristics of the RF links into account, we derive closed-form expressions for the network outage probability. We also evaluate the effect of various parameters such as power amplifiers efficiency, number of antennas as well as different coherence times of the RF and the FSO links on the system performance. Finally, we present mappings between the performance of RF-FSO multi-hop networks and the ones using only the RF- or the FSO-based communication, in the sense that with appropriate parameter settings the same outage probability is achieved in these setups. The results show the efficiency of the RF-FSO setups in different conditions. Moreover, the HARQ can effectively improve the outage probability/energy efficiency, and compensate the effect of hardware impairments in RF-FSO networks. For common parameter settings of the RF-FSO dual-hop networks, outage probability $10^{-4}$ and code rate $3$ nats-per-channel-use, the implementation of HARQ with a maximum of $2$ and $3$ retransmissions reduces the required power, compared to the cases with no HARQ, by $13$ and $17$ dB, respectively.
\end{abstract}



%
\IEEEpeerreviewmaketitle
\vspace{-2mm}
\section{Introduction}
To address the demands on the next generation of wireless networks, a combination of different techniques are considered among which free-space optical (FSO) communication is very promising  \cite{6887284}. FSO systems provide fiber-like data rates through the atmosphere using lasers or light emitting diodes (LEDs). Thus, FSO can be used for a wide range of applications such as last-mile access, back-hauling and multi-hop networks. In the radio frequency (RF) domain, on the other hand, it has been recently concentrated on millimeter wave (MMW) communication as a key enabler to obtain sufficiently large bandwidths so that it is possible to achieve data rates comparable to those in the FSO links. In this way, the combination of FSO and MMW-based RF links is considered as a powerful candidate for high-rate reliable communication in, e.g., vehicle- and infrastructure-to-infrastructure networks. This is particularly interesting because, as we show in the following, with proper parameter settings the outage probability of the RF-FSO networks can be mapped to the ones in the cases with only the RF- or the FSO-based communication.

The RF-FSO related works can be divided into two categories. The first group are papers on single-hop setups where the link reliability is improved via the joint implementation of RF and FSO systems. Here, either the RF and the FSO links are considered as separate links and the RF link acts as a backup when the FSO link is down, e.g., \cite{6887284,6364576,6400459}, or the links are combined to improve the system performance \cite{6503564,5342330,5427418}.  Also, the implementation of hybrid automatic repeat request (HARQ) in RF-FSO links has been considered in \cite{5427418,6692504,7445896}.

The second group are the papers studying the performance of multi-hop RF-FSO networks. For instance, \cite{6775014} analyzes decode-and-forward techniques in multiuser relay networks using RF-FSO. Then, \cite{6831655,6866170} study RF-FSO based relaying with an RF source-relay link and an FSO  or RF-FSO relay-destination link. Also, considering Rayleigh fading conditions for the RF link and amplify-and-forward relaying technique, \cite{6678140,5999707} derive the end-to-end error probability of the RF-FSO based setups and compare the system performance with RF-based relay networks, respectively. Finally, the impact of pointing errors on the performance of dual-hop RF-FSO systems is studied in \cite{6512100,7055847}.

In this paper, we analyze the performance of multi-hop RF-FSO systems from an information theoretic point of view.
Considering the MMW characteristics of the RF links and heterodyne detection technique in the FSO links, we derive closed-form expressions for the system outage probability  (Lemmas 1-3). Our results are obtained for the decode-and-forward relaying approach in different cases with and without HARQ. Specifically, we show the HARQ as an effective technique to compensate for the imperfect properties of the RF-FSO system and to improve the network reliability/energy efficiency. Finally, we present mappings between the performance of RF- and FSO-based hops (Corollary 1),
and analyze the effect of various parameters such as the power amplifiers (PAs) efficiency, different coherence times of the links and number of antennas on the network outage probability.

As opposed to \cite{6887284,6364576,6400459,6503564,5342330,5427418,6692504,7445896}, we consider multi-hop systems. Also, our paper is different from \cite{6887284,6364576,6400459,6503564,5342330,5427418,6692504,7445896,6831655,6866170,6678140,5999707,6512100,7055847,6775014} because our  analytical/numerical results on the outage probability as well as our discussions on the effect of imperfect PAs and HARQ have not been presented before. The differences in the problem formulation and the channel model makes our analytical/numerical results as well as our conclusions completely different from the ones in the literature.

Our results show that there are mappings between the performance of RF-FSO multi-hop networks and the ones using only the RF- or the FSO-based communication, in the sense that with proper scaling of the channel parameters the same outage probability is achieved in these setups (Corollary 1).
While the outage probability is sensitive to the number of RF-based transmit antennas for short codewords, the outage probability reduction due to increasing the number of antennas is negligible for the cases with long codewords.
The PAs efficiency affects the network outage probability considerably. However, the HARQ protocols can effectively  compensate the effect of hardware impairments. Finally, the HARQ improves the outage probability/energy efficiency significantly. For instance, consider common parameter settings of the RF-FSO dual-hop networks, outage probability $10^{-4}$ and code rate $3$ nats-per-channel-use (npcu). Then, compared to the cases with open-loop communication, the implementation of HARQ with a maximum of $2$ and $3$ retransmissions reduces the required power by $13$ and $17$ dB, respectively.
\vspace{-2mm}
\section{System Model}
Consider a $T^\text{total}$-hop RF-FSO system, with $T$ RF-based hops and $\tilde T=T^\text{total}-T$ FSO-based hops. As seen in the following, the outage probability is independent of the order of the hops. Thus, we do not need to specify the order of the RF- and FSO-based hops. The $i$-th, $i=1,\ldots,T,$ RF-based hop uses a multiple-input-single-output (MISO) setup with  $N_{i}$ transmit antennas.
We define the channel gains as $g_{i}^{j_i}\doteq|h_{i}^{j_i}|^2,i=1,\ldots, T,j_i=1,\ldots,N_i,$ where $h_i^{j_i}$ is the complex fading coefficients of the channel between the $j_i$-th antenna in the $i$-th hop and its corresponding receive antenna.

Here, we present the analytical results for the Rician channel model of the RF-based hops, which is an appropriate model for near line-of-sight conditions and has been well established for different millimeter wave-based applications, e.g., \cite{mmwrician1,mmwrician2,mmwrician3}. Let us denote the probability density function (PDF) and the cumulative distribution function (CDF) of a random variable $X$ by $f_X(\cdot)$ and $F_X(\cdot)$, respectively. With a Rician model, the channel gain ${g_i^{j_i},\forall i,j_i,}$ follows the PDF
\vspace{-2mm}
\begin{align}\label{eq:eqRicianpdf}
&f_{g_i^{j_i}}(x)=\frac{(K_i+1)e^{-K_i}}{\Omega_i}\times \nonumber\\&e^{-\frac{(K_i+1)x}{\Omega_i}}I_0\left(2\sqrt{\frac{K_i(K_i+1)x}{\Omega_i}} \right ),\forall i,j_i,
\end{align}
where $K_i$ and $\Omega_i$ denote the fading parameters in the $i$-th hop and $I_n(\cdot)$ is the $n$-th order modified Bessel function of the first kind.
Also, the sum channel gain $G_i=\sum_{j_i=1}^{N_i}{g_i^{j_i}}$ follows
\vspace{-2mm}
\begin{align}\label{eq:eqRiciansumpdf}
f_{G_i}(x)&=\frac{(K_i+1)e^{-K_iN_i}}{\Omega_i}\left(\frac{(K_i+1)x}{K_iN_i\Omega_i}\right)^{\frac{N_i-1}{2}}\times\nonumber\\&e^{-\frac{(K_i+1)x}{\Omega_i}}I_{N_i-1}\left(2\sqrt{\frac{K_i(K_i+1)N_ix}{\Omega_i}} \right ),\forall i.
\end{align}

Finally, to take the non-ideal hardware into account, we consider the state-of-the-art model for the PA efficiency where the output power at each antenna of the $i$-th hop is determined according to
  \cite[eq. (3)]{6725577}, \cite[eq. (1)]{7104158}
  \vspace{-2mm}
\begin{align}\label{eq:ampmodeldaniel}
&\frac{P_i}{P_i^{\text{cons}}}=\epsilon_i\left(\frac{P_i}{P_i^{\text{max}}}\right)^\vartheta_i\Rightarrow  P_i=\sqrt[1-\vartheta_i]{\frac{\epsilon_i P_i^{\text{cons}}}{(P_i^{\text{max}})^\vartheta_i}},\forall i.
\end{align}
Here, $ P_i, P_i^{\text{max}}$ and $P_i^{\text{cons}},\forall i,$ are the output, the maximum output and the consumed power in each antenna of the $i$-th hop, respectively, $\epsilon_i\in [0,1]$ denotes the maximum power efficiency achieved at $P_i=P_i^{\text{max}}$ and $\vartheta_i\in [0,1]$ is a parameter depending on the PA classes.

The FSO links, on the other hand, are assumed to have single transmit/receive terminals. Reviewing the literature and depending on the channel condition, the FSO link may follow different distributions. Here, we present the results for the cases with exponential and Gamma-Gamma distributions of the FSO links. For the exponential distribution of the $i$-th FSO hop, the channel gain $\tilde G_i$ follows
\vspace{-2mm}
\begin{align}\label{eq:Eqexpfsopdf}
f_{\tilde G_i}(x)=\lambda_i e^{-\lambda_i x}, \forall i,
\end{align}
 with $\lambda_i$ being the long-term channel coefficient of the $i$-th, $i=1,\ldots,\tilde T,$ hop. Moreover, with the Gamma-Gamma distribution we have
\vspace{-2mm}
\begin{align}\label{eq:eqpdfgammagamma}
f_{\tilde G_i}(x)=\frac{2(a_ib_i)^{\frac{a_i+b_i}{2}}}{\Gamma(a_i)\Gamma(b_i)}x^{\frac{a_i+b_i}{2}-1}\mathcal{K}_{a_i-b_i}\left(2\sqrt{a_ib_ix}\right),\forall i.
\end{align}
Here, $\mathcal{K}_n(\cdot)$ denotes the modified Bessel function of the second kind of order $n$ and $\Gamma(x)=\int_0^\infty{u^{x-1}e^{-u}\text{d}u}$ is the Gamma function. Also, $a_i$ and $b_i, i=1,\ldots,\tilde T,$ are the distribution shaping parameters which can be expressed as functions of Rytov variance, e.g., \cite{7445896}.
\vspace{-2mm}
\subsection{Data Transmission Model}
We consider the decode-and-forward technique where at each hop the received message is decoded and re-encoded, if it is correctly decoded. Thus, the message is successfully received by the destination if it is correctly decoded in all hops. Otherwise, outage occurs.  As the most promising HARQ approach leading to lowest outage probability \cite{throughputdef}, we consider the incremental redundancy (INR) HARQ with a maximum of $M_i$ retransmissions in the $i$-th, $i=1,\ldots,T^\text{total},$ hop. Using INR HARQ with a maximum of  $M_i$ retransmissions, $q_i$ information nats are encoded into a codeword of length $M_iL$ channel uses. Then, the codeword is divided into $M_i$ sub-codewords of length $L$ channel uses which are sent in the successive retransmissions. Thus, the equivalent data rate at the end of round $m$ is $\frac{q_i}{mL}=\frac{R_i}{m}$ npcu where $R_i=\frac{q_i}{L}$ denotes the initial code rate in the $i$-th hop. In each round, the data is decoded based on all sub-codewords received up to the end of that round. The retransmission continues until the message is correctly decoded or the maximum permitted transmission round is reached. Finally, note that setting $M_i=1,\forall  i,$ represents the cases without HARQ, i.e., open-loop communication.
\vspace{-5mm}
\section{Analytical results}
Considering the decode-and-forward approach and because independent channel realizations are experienced in different hops, the system outage probability is given by
\vspace{-3mm}
\begin{align}\label{eq:Eqtotaloutage}
\Pr\left(\text{Outage}\right)=1-\prod_{i=1}^{T}{(1-\phi_i)}\prod_{i=1}^{\tilde T}{(1-\tilde \phi_i)}.
\end{align}
Here, $\phi_i$ and $\tilde \phi_i$ denote the outage probability in the $i$-th RF- and FSO-based hops, respectively. Thus, to analyze the outage probability, we need to find  $\phi_i$ and $\tilde \phi_i,\forall i$. Following the same procedure as in, e.g.,  \cite{throughputdef} and using the properties of the imperfect PAs (\ref{eq:ampmodeldaniel}), the outage probability of the $i$-th RF- and FSO-based hops are obtained as
\vspace{-3mm}
\begin{align}\label{eq:EqtotaloutageRF}
&\phi_i=\Pr\Bigg(\frac{1}{M_iC_i}\times\nonumber\\&\sum_{m=1}^{M_i}\sum_{c=(m-1)C_i+1}^{mC_i}\log\left(1+\sqrt[1-\vartheta_i]{\frac{\epsilon_i P_i^{\text{cons}}}{(P_i^{\text{max}})^\vartheta_i}}G_i(c)\right)\le\frac{R_i}{M_i}\Bigg),
\end{align}
and
\vspace{-5mm}
\begin{align}\label{eq:EqtotaloutageFSO}
&\tilde \phi_i=\Pr\Bigg(\frac{1}{M_i\tilde C_i}\times\nonumber\\&\sum_{m=1}^{M_i}\sum_{c=(m-1)\tilde C_i+1}^{m\tilde C_i}\log\left(1+\tilde P_i\tilde G_i(c)\right)\le\frac{R_i}{M_i}\Bigg),
\end{align}
respectively. Here, $\tilde P_i$ represents the transmission power in the $i$-th, $i=1,\ldots,\tilde T,$ FSO-based  hop and we have considered heterodyne detection technique in (\ref{eq:EqtotaloutageFSO}). Also, $C_i, i=1,\ldots,T,$ and $\tilde C_i, i=1,\ldots,\tilde T,$ represent the number of channel realizations experienced in each HARQ-based transmission round of the $i$-th RF- and FSO-based hops, respectively (for simplicity, we present the results for the cases with normalized symbol rates. Using the same approach as in \cite{5342330}, it is straightforward to represent the results with different symbol rates of the links). In the sequel, we present closed-form expressions for (\ref{eq:EqtotaloutageRF})-(\ref{eq:EqtotaloutageFSO}), and, consequently, (\ref{eq:Eqtotaloutage}). 
Here, we concentrate on the cases with long codewords where multiple channel realizations are experienced during data transmission in each hop, i.e., $C_i$ and $\tilde C_i,\forall i,$ are assumed to be large. For performance comparison in the cases with short and long codewords, see Fig. 4. Indeed, to find the expressions, we need to implement approximation techniques. However, as seen in Section IV, the analytical results mimic the numerical results with high accuracy.

To approximate (\ref{eq:EqtotaloutageRF}), we first represent an approximation for the PDF of the sum channel gain $G_i,\forall i,$ as follows.

\textbf{\emph{Lemma 1:}} For moderate/large number of antennas, which is of interest in MMW communication,  the sum gain $G_i,\forall i,$ is approximated by an equivalent Gaussian random variable $\mathcal{Z}_i\sim\mathcal{N}(N_i\zeta_i,N_i\nu_i^2)$ with $\zeta_i=\mathcal{S}_i(2)$, $\nu_i^2=\mathcal{S}_i(4)-\mathcal{S}_i(2)^2$ and $\mathcal{S}_i(n)\doteq\left(\frac{\Omega_i}{K_i+1}\right)^\frac{n}{2}\Gamma\left(1+\frac{n}{2}\right)\mathcal{L}_{\frac{n}{2}}\left(-K_i\right)$. Here, $\mathcal{L}_{n}(x)=\frac{e^x}{n!}\frac{\mathrm{d}^n }{\mathrm{d} x^n}\left(e^{-x}x^n\right)$ denotes the Laguerre polynomial of the $n$-th order and $K_i,\Omega_i$ are the fading parameters as defined in (\ref{eq:eqRicianpdf}).

\begin{proof}
Using central limit Theorem (CLT) for moderate/large number of antennas, the random variable  $G_i=\sum_{j_i=1}^{N_i}{g_i^{j_i}}$ is approximated by the Gaussian random variable $\mathcal{Z}_i\sim\mathcal{N}(N_i\zeta_i,N_i\nu_i^2)$. Here, from (\ref{eq:eqRicianpdf}), $\zeta_i$ and  $\nu_i^2$ are found as
\vspace{-2mm}
\begin{align}\label{eq:eqmeanCLTmmw}
&\zeta_i=\int_0^\infty{xf_{{g_i^{j_i}}}(x)\text{d}x}=\frac{(K_i+1)e^{-K_i}}{\Omega_i}\times\nonumber\\&\int_0^\infty{xe^{-\frac{(K_i+1)x}{\Omega_i}}I_0\left(2\sqrt{\frac{K_i(K_i+1)x}{\Omega_i}} \right )\text{d}x},
\end{align}
and
\vspace{-2mm}
\begin{align}\label{eq:eqmeanCLTmmw}
&\nu_i^2=\rho_i-\zeta_i^2,\nonumber\\&
\rho_i=\int_0^\infty{x^2f_{{g_i^{j_i}}}(x)\text{d}x}\nonumber\\&=\frac{(K_i+1)e^{-K_i}}{\Omega_i}\int_0^\infty{x^2e^{-\frac{(K_i+1)x}{\Omega_i}}I_0\left(2\sqrt{\frac{K_i(K_i+1)x}{\Omega_i}} \right )\text{d}x},
\end{align}
respectively. Then, using the variable transform $t=\sqrt{x}$, some manipulations and the properties of the Bessel function $\frac{1}{b^2}\int_0^\infty{x^{n+1}e^{-\frac{x^2+c^2}{2b^2}}I_0\left(\frac{cx}{b^2}\right)\text{d}x}=b^n2^\frac{n}{2}\Gamma\left(1+\frac{n}{2}\right)\mathcal{L}\left(\frac{-c^2}{2b^2}\right),\forall c,b,n$, the mean and the variance $\zeta_i, \nu_i^2$ are determined as stated in the lemma.
\end{proof}

\textbf{\emph{Lemma 2:}} The outage probability (\ref{eq:EqtotaloutageRF}) is approximated by (\ref{eq:eqlemma3aa}) with $\hat\mu_i$ and $\hat\sigma_i^2$ given in (\ref{eq:eqtheoremmu2})-(\ref{eq:eqtheoremsigma2}), respectively.

\begin{proof}
Replacing the random variable $\frac{1}{M_iC_i}\sum_{m=1}^{M_i}\sum_{c=(m-1)C_i+1}^{mC_i}\log\left(1+\sqrt[1-\vartheta_i]{\frac{\epsilon_i P_i^{\text{cons}}}{(P_i^{\text{max}})^\vartheta_i}}G_i(c)\right)$ by its equivalent CLT-based Gaussian random variable $\mathcal{U}_i\sim\mathcal{N}\left(\hat\mu_i,\frac{1}{M_iC_i}\hat\sigma_i^2\right)$,  the probability (\ref{eq:EqtotaloutageRF}) is approximated by
\begin{align}\label{eq:eqapproxlemma3}
\phi_i\simeq\Pr\left(\mathcal{U}_i\le\frac{R_i}{M_i}\right), \mathcal{U}_i\sim\mathcal{N}\left(\hat\mu_i,\frac{1}{M_iC_i}\hat\sigma_i^2\right),
\end{align}
where
\begin{align}\label{eq:eqtheoremmu2}
\hat \mu_i&=\int_0^\infty{\log\left(1+\sqrt[1-\vartheta_i]{\frac{\epsilon_i P_i^{\text{cons}}}{(P_i^{\text{max}})^\vartheta_i}}x\right)}f_{G_i}(x)\text{d}x
\nonumber\\&\mathop  \simeq \limits^{(a)}
\int_0^\infty{\mathcal{Y}_i(x)f_{\mathcal{Z}_i}(x)\text{d}x}
\nonumber\\&=
\mathcal{Q}\left(\sqrt[1-\vartheta_i]{\frac{\epsilon_i P_i^{\text{cons}}}{(P_i^{\text{max}})^\vartheta_i}},0,N_i\zeta_i,N_i\nu_i^2,s_i\right)\nonumber\\&-\mathcal{Q}\left(\sqrt[1-\vartheta_i]{\frac{\epsilon_i P_i^{\text{cons}}}{(P_i^{\text{max}})^\vartheta_i}},0,N_i\zeta_i,N_i\nu_i^2,0\right)\nonumber\\&+\mathcal{Q}\left(r_i,\theta-r_id_i,N_i\zeta_i,N_i\nu_i^2,\infty\right)
\nonumber\\&-\mathcal{Q}\left(r_i,\theta-r_id_i,N_i\zeta_i,N_i\nu_i^2,s_i\right),
\nonumber\\&
\mathcal{Q}(a_1,a_2,a_3,a_4,x)\nonumber\\&\doteq-\frac{a_1a_3+a_2}{2}\text{erf}\left(\frac{a_3-x}{\sqrt{2a_4}}\right)-\frac{a_4}{2\pi}a_1e^{-\frac{(a_3-x)^2}{2a_4}},
\end{align}
and
\begin{align}
&\hat\sigma_i^2=\hat\gamma_i-\hat\mu_i^2,\nonumber\\&
=\hat \gamma_i=\int_0^\infty{\log^2\left(1+\sqrt[1-\vartheta_i]{\frac{\epsilon_i P_i^{\text{cons}}}{(P_i^{\text{max}})^\vartheta_i}}x\right)}f_{G_i}(x)\text{d}x
\nonumber\\&
\mathop  \simeq \limits^{(b)}
\int_0^\infty{\mathcal{Y}_i^2(x)f_{\mathcal{Z}_i}(x)\text{d}x}
\nonumber\\&
= \mathcal{T}\left(\sqrt[1-\vartheta_i]{\frac{\epsilon_i P_i^{\text{cons}}}{(P_i^{\text{max}})^\vartheta_i}},0,N_i\zeta_i,N_i\nu_i^2,s_i\right)\nonumber
\end{align}
\begin{align}\label{eq:eqtheoremsigma2}
&-\mathcal{T}\left(\sqrt[1-\vartheta_i]{\frac{\epsilon_i P_i^{\text{cons}}}{(P_i^{\text{max}})^\vartheta_i}},0,N_i\zeta_i,N_i\nu_i^2,0\right)\nonumber\\&+\mathcal{T}\left(r_i,\theta-r_id_i,N_i\zeta_i,N_i\nu_i^2,\infty\right)
\nonumber\\&-\mathcal{T}\left(r_i,\theta-r_id_i,N_i\zeta_i,N_i\nu_i^2,s_i\right),
\nonumber\\&
\mathcal{T}\left(a_1,a_2,a_3,a_4,x\right)\doteq\frac{1}{2\sqrt{2\pi}}e^{-\frac{x^2+a_3^2}{2a_4}}\Bigg(\text{erf}\left(\frac{x-a_3}{\sqrt{2a_4}}\right)\nonumber\\&-2\sqrt{a_4}a_1e^{\frac{a_3x}{a_4}}(a_1(a_3+x)+2a_2)\nonumber\\&+\sqrt{2\pi}e^{\frac{x^2+a_3^2}{2a_4}}\left(a_1^2(a_3^2+a_4)+2a_1a_2a_3+a_2^2\right)
\Bigg).
\end{align}
Here, $(a)$ and $(b)$ in (\ref{eq:eqtheoremmu2}) and (\ref{eq:eqtheoremsigma2}) come from approximating $f_{G_i}(x)$ by $f_{\mathcal{Z}_i}(x)$ defined in Lemma 1 and the approximation $\log\left(1+\sqrt[1-\vartheta_i]{\frac{\epsilon_i P_i^{\text{cons}}}{(P_i^{\text{max}})^\vartheta_i}}x\right)\simeq \mathcal{Y}_i(x)$ where
%
%
%

\begin{align}
&\mathcal{Y}_i(x)=\left\{\begin{matrix}
{\sqrt[1-\vartheta_i]{\frac{\epsilon_i P_i^{\text{cons}}}{(P_i^{\text{max}})^\vartheta_i}}}x,\,\,\,\,\,\,\,\,\,\,\,\,\,\,\,\,\,\,\,\,\,\,\,\,\,\,\,\,\,\,\,\,\,\,\,\,\,\,\,\, x\in\left[0,s_i\right]\,\,\,\,\,\,\,\,\,\,\,\,\,\,\,\,\,\,\,\,\,\,\,\,\,\,\,\,\,\,\,\,\,\,\\
\theta+r_i\left(x-d_i \right ), \,\,\,\,\,\,\,\,\,\,\,\,\,\,\,\,\,\,\,\,\,\,\,\,\,\,\,\,\,\,\,\,\,\,  x> s_i,\,\,\,\,\,\,\,\,\,\,\,\,\,\,\,\,\,\,\,\,\,\,\,\,\,\,\,\,\,\,\,\,\,\,\,\,\,\,\,\,\,
\end{matrix}\right.,
\nonumber\\&
s_i=\frac{\theta }{\sqrt[1-\vartheta_i]{\frac{\epsilon_i P_i^{\text{cons}}}{(P_i^{\text{max}})^\vartheta_i}}\left(1-e^{-\theta}\right)}-\frac{1}{\sqrt[1-\vartheta_i]{\frac{\epsilon_i P_i^{\text{cons}}}{(P_i^{\text{max}})^\vartheta_i}}},
\nonumber\\&
r_i=\sqrt[1-\vartheta_i]{\frac{\epsilon_i P_i^{\text{cons}}}{(P_i^{\text{max}})^\vartheta_i}}e^{-\theta}
,
d_i=\frac{e^\theta-1}{\sqrt[1-\vartheta_i]{\frac{\epsilon_i P_i^{\text{cons}}}{(P_i^{\text{max}})^\vartheta_i}}}.
\end{align}
Then, using the CDF of Gaussian random variables and the error function $\text{erf}(x)=\frac{2}{\sqrt{\pi}}\int_0^x{e^{-t^2}\text{d}t}$, (\ref{eq:eqapproxlemma3}) is determined as
\begin{align}\label{eq:eqlemma3aa}
\phi_i&\simeq
\frac{1}{2}\left(1+\text{erf}\left(\frac{\sqrt{M_iC_i}\left(\frac{R_i}{M_i}-\hat\mu_i\right)}{\sqrt{2\hat\sigma_i^2}}\right)\right),\forall \theta>0.
\end{align}
Note that, in (\ref{eq:eqlemma3aa}), $\theta>0$ is an arbitrary parameter and, based on our simulation, accurate approximations are obtained for a broad range of $\theta>0.$
\end{proof}
Finally, the following lemma represents the outage probability of the FSO-based hops.

\textbf{\emph{Lemma 3:}} The outage probability of the FSO-based hop, i.e., (\ref{eq:EqtotaloutageFSO}), is approximately given by
\begin{align}\label{eq:eqlemma4aa}
\tilde\phi_i&\simeq
\frac{1}{2}\left(1+\text{erf}\left(\frac{\sqrt{M_i\tilde C_i}\left(\frac{R_i}{M_i}-\tilde\mu_i\right)}{\sqrt{2\tilde\sigma_i^2}}\right)\right).
\end{align}
Here, $\tilde \mu_i$ and $\tilde\sigma_i^2$ are given by (\ref{eq:mueq})-(\ref{eq:sigmaeq}) and \cite[eq. 43-44]{7445896} for the exponential and the Gamma-Gamma distributions of the FSO links, respectively.
\begin{proof}
Using CLT, the random variable $\frac{1}{M_i\tilde C_i}\sum_{m=1}^{M_i}\sum_{c=(m-1)\tilde C_i+1}^{mC_i}\log\left(1+\tilde P_i\tilde G_i(c)\right)$ is replaced by its equivalent Gaussian random variable $\mathcal{R}_i\sim\mathcal{N}(\tilde\mu_i,\frac{1}{M_i\tilde C_i}\tilde\sigma_i^2)$,  where for the exponential distribution of the FSO link we have
\vspace{-0mm}
\begin{align}\label{eq:mueq}
\tilde \mu_i&=\int_0^\infty{f_{{\tilde G_i}}(x)\log(1+\tilde P_ix)\text{d}x}\nonumber\\&\mathop  = \limits^{(c)} \tilde P_i\int_0^\infty{\frac{1-F_{\tilde G_i}(x)}{1+\tilde P_ix}\text{d}x}=-e^{\frac{\tilde \lambda_i }{\tilde P_i}}\text{Ei}\left(-{\frac{\lambda_i }{\tilde P_i}}\right),
\end{align}
and
\vspace{-2mm}
\begin{align}\label{eq:sigmaeq}
&\tilde \sigma_i^2=\tilde \rho_i-\tilde \mu_i^2,\nonumber\\&
\tilde \rho_i=E\{\log(1+\tilde P_i\tilde G_i)^2\}=\int_0^\infty{f_{{\tilde  G_i}}(x)\log^2(1+\tilde P_ix)\text{d}x} \nonumber\\& \mathop  = \limits^{(d)}{2\tilde P_i}\int_0^\infty{\frac{e^{- \lambda_i x}}{1+\tilde P_ix}\log(1+\tilde P_ix)\text{d}x}\mathop  = \limits^{(e)} \mathcal{H}_i\left(\infty\right)-\mathcal{H}_i\left(1\right),\nonumber\\&
\mathcal{H}_i(x)=2e^{\frac{\lambda_i}{\tilde P_i}}\bigg(\frac{\lambda_i}{\tilde P_i}x\prescript{}{3}F_{3}\left(1,1,1;2,2,2;-\frac{\lambda_i x}{\tilde P_i}\right)\nonumber\\&+\frac{1}{2}\log(x)\bigg(-2\left(\log\left(\frac{\lambda_i}{\tilde P_i}x\right)+\mathcal{E}\right)\nonumber\\&-2\Gamma\left(0,\frac{\lambda_i}{\tilde P_i}x\right)+\log(x)\bigg)\bigg).
\end{align}
Here, $\text{Ei}(x)=\int_x^\infty{\frac{e^{-t}\text{d}t}{t}}$ represents the exponential integral function. Moreover,  $(c)$ and $(d)$ are obtained by partial integration. Then, denoting the Euler constant by $\mathcal{E}$, $(e)$ comes from variable transform $1+\tilde P_ix=t,$ some manipulations as well as the definition of Gamma incomplete function $\Gamma(s,x)=\int_x^\infty{t^{s-1}e^{-t}\text{d}t}$ and the generalized hypergeometric function $\prescript{}{n_1}F_{n_2}(\cdot).$ For the Gamma-Gamma distribution, on the other hand, the PDF $f_{\tilde G_i}$ in (\ref{eq:mueq})-(\ref{eq:sigmaeq}) is replaced by (\ref{eq:eqpdfgammagamma}) and the mean and variance are calculated by \cite[eq. 43]{7445896} and \cite[eq. 44]{7445896}, respectively. In this way, following the same arguments as in Lemma 2, the outage probability of the FSO-based hops is approximated by (\ref{eq:eqlemma4aa}).
\end{proof}
Lemmas 1-3 lead to the following corollary statement about the performance of multi-hop RF-FSO systems.

\textbf{\emph{Corollary 1:}} With long codewords, there are mappings between the performance of FSO- and RF-based hops, in the sense that, with proper parameter settings, the same outage probability is achieved in these hops.
\begin{proof}
The proof comes from Lemmas 1-3 where for different hops the outage probability is given by the CDF of Gaussian random variables. Thus, with parameter settings of  $(\hat\mu_i,\hat\sigma_i)$  and $(\tilde \mu_i,\tilde \sigma_i)$ in Lemmas 2-3, the same outage probability is achieved in these hops. In this way, the performance of RF-FSO based multi-hop networks can be mapped to the ones using only the RF- or the FSO-based communication.
\end{proof}

\vspace{-1mm}
\section{Simulation Results}
Throughout the paper, we presented different approximation techniques. The verification of these results is demonstrated in Fig. 1 and, as seen in the sequel, the analytical results follow the simulations with high accuracy. Then, to avoid too much information in each figure, Figs. 2-4 report only the simulation results. Note that in all simulations we have checked the results with the ones obtained analytically and they match tightly.

The simulation results are presented for homogenous setups. That is, different RF-based hops follow the same long-term fading parameters $K_i,\omega_i,\forall i,$ in (\ref{eq:eqRicianpdf})-(\ref{eq:eqRiciansumpdf}), and the FSO-based hops also experience the same long-term channel parameters, i.e., $\lambda_i,a_i$ and $b_i$ in (\ref{eq:Eqexpfsopdf})-(\ref{eq:eqpdfgammagamma}). Also, we consider $M_i=M_j$ and $R_i=R_j,\forall i,j=1,\ldots, T.$ In all figures, we set  $\tilde P_i=N_iP_i^\text{cons}$ such that the total consumed power at different hops are the same. Then, using (\ref{eq:ampmodeldaniel}), one can determine the output of the RF-based antennas. Also, because the noise variances are set to 1, $\tilde P_i$ (in dB, $10\log_{10}\tilde P_i$) is referred to the SNR as well. In Figs. 1 and 4, we assume an ideal PA. The effect of non-ideal PAs is verified in Figs. 2 and 3. With non-ideal PAs, we consider the state-of-the-art parameter settings $\vartheta_i=0.5, \epsilon_i=0.75, P_i^\text{max}=25 \text{ dB},\forall i,$ unless otherwise stated \cite{6725577,7104158}. The parameters of the Rician RF PDF (\ref{eq:eqRicianpdf}) are set to $\omega_i=1, K_i=0.01, \forall i,$ leading to unit mean and variance of the channel gain distribution $f_{g_i^{j_i}}(x),\forall i,j_i$. With the exponential distribution of the FSO-based hops, we consider $f_{\tilde G_i}(x)=\lambda_i e^{-\lambda_i x}$ with $\lambda_i=1,\forall i.$ Also, for the Gamma-Gamma distribution we set $f_{\tilde G_i}(x)=\frac{2(a_ib_i)^{\frac{a_i+b_i}{2}}}{\Gamma(a_i)\Gamma(b_i)}x^{\frac{a_i+b_i}{2}-1}\mathcal{K}_{a_i-b_i}\left(2\sqrt{a_ib_ix}\right),$ $a_i=4.3939, b_i=2.5636, \forall i,$ which corresponds to Rytov variance 1 \cite{7445896}. 

\begin{figure}
\centering
  \includegraphics[width=0.8\columnwidth]{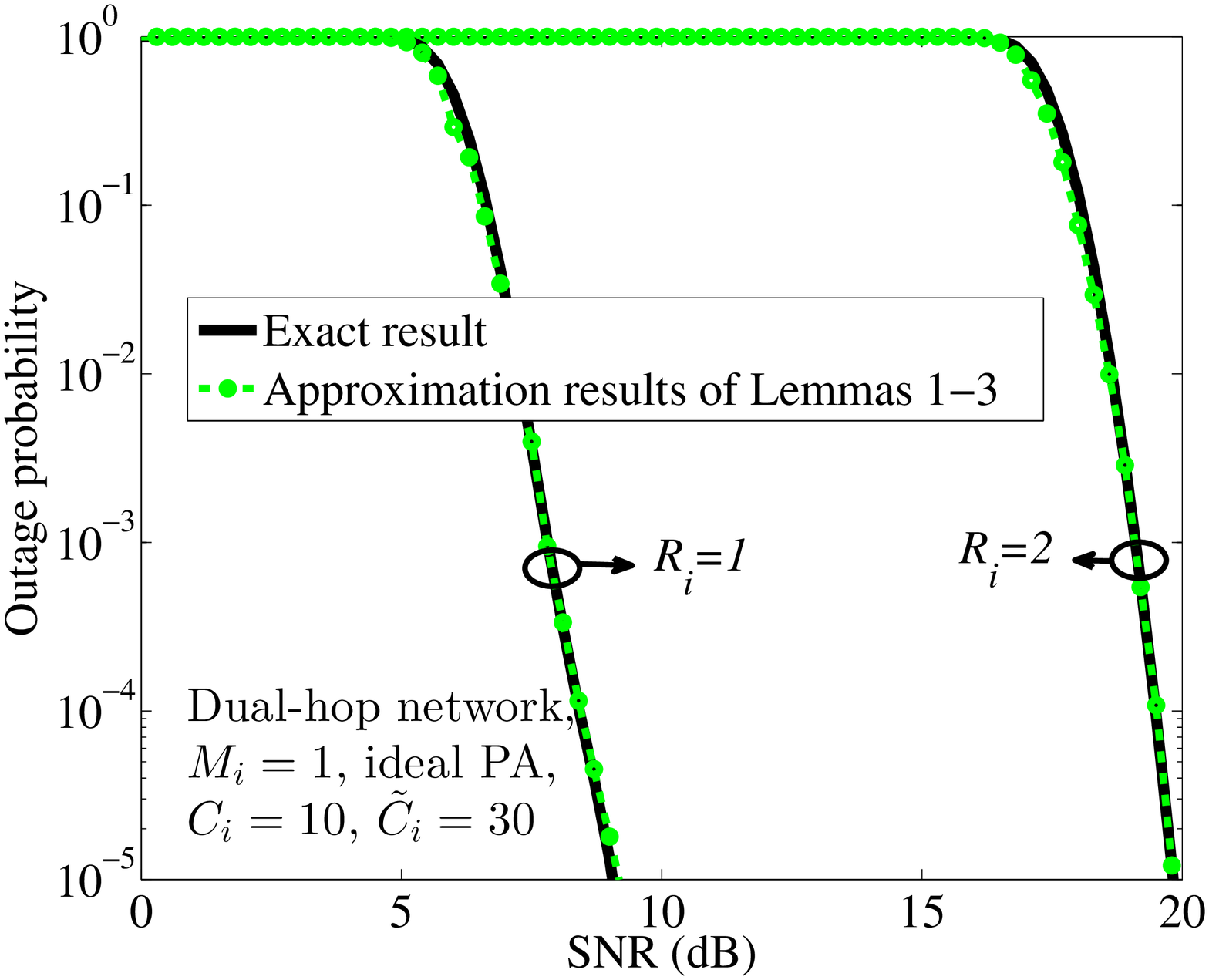}\\\vspace{-3mm}
\caption{On the tightness of the results in Lemmas 1-3. Ideal PA, dual-hop network, $M_i=1, R_i=1,2, C_i=10, \tilde C_i=30, T=1,\tilde T=1, N_i=20.$}\vspace{-3mm}\label{figure111}
\end{figure}

\begin{figure}
\centering
  \includegraphics[width=0.8\columnwidth]{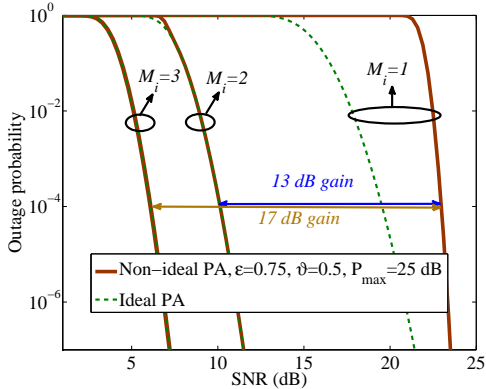}\\\vspace{-3mm}
\caption{Outage probability of dual-hop RF-FSO network for different PA models and maximum number of retransmissions, $M_i,\forall i$. Exponential distribution of the FSO link, $T=1, \tilde T=1, C_i=10, \tilde C_i=20, R_i=3$ npcu, and  $N_i=60, \forall i.$}\vspace{-3mm}\label{figure111}
\end{figure}

The simulation results are presented in different parts as follows.

\emph{On the approximation approaches of Lemmas 1-3:} Considering an ideal PA and $M_i=1, R_i=2$ npcu, and $C_i=10, \forall i,$ Fig. 1 verifies the tightness of the approximation schemes of Lemmas 1-3. Particularly, we plot the outage probability of a dual-hop RF-FSO setup versus the SNR.  Here, we set $M_i=1, R_i=1,2$ npcu, and $C_i=10, \tilde C_i=30, N_i=20, T=1, \tilde T=1 \forall i,$ and the results are presented for the cases with ideal PAs at the RF-based hops. As it is observed, the analytical results of Lemmas 1-3 mimic the exact results with very high accuracy. As a result, (\ref{eq:eqlemma3aa}) and (\ref{eq:eqlemma4aa}) can be effectively used to analyze the data transmission efficiency of the RF-FSO multi-hop networks, as well as the multi-hop networks with only the RF- or the FSO-based communication.

\emph{On the effect of HARQ and imperfect PAs:} Figure 2 shows the outage probability of a dual-hop RF-FSO network for different maximum numbers of HARQ-based retransmission rounds $M_i,\forall i.$ Also, the figure compares the system performance in the cases with ideal and non-ideal PAs. Here, the results are presented for the exponential distribution of the FSO link, $T=1, \tilde T=1, C_i=10, \tilde C_i=20, R_i=3$ npcu, and  $N_i=60, \forall i.$ As demonstrated, with no HARQ, the efficiency of the RF-based PAs affects the system performance considerably. For instance, with the parameter settings of the figure and outage probability $10^{-4},$ the PAs inefficiency increases the required power by $3.5$ dB. On the other hand, the HARQ can effectively compensate the effect of imperfect PAs, and the difference between the outage probability of the cases with ideal and non-ideal PAs is negligible for $M>1.$ 
Finally, the HARQ improves the energy efficiency significantly. As  an example, consider the outage probability $10^{-4}$, an ideal PA and the parameter settings of Fig. 2. Then, compared to the open-loop communication, i.e., $M_i=1$, the implementation of HARQ with a maximum of 2 and 3 retransmissions reduces the required power by 13 and 17 dB, respectively.

\emph{System performance with different numbers of hops:} In Fig. 3, we show the outage probability in the cases with different numbers of RF- and FSO-based hops, i.e., $T, \tilde T$. As expected, the outage probability increases with the number of hops. However, the outage probability increment is negligible particularly at high SNRs because the data is correctly decoded with high probability in different hops as the SNR increases. Finally, the figure indicates that the outage probability of the RF-FSO based multi-hop network is not sensitive to the distribution of the FSO-based hops at low SNRs. Intuitively, this is because at low SNRs and with the parameter settings of the figure the outage event mostly occurs in the RF-based hops. However, at high SNRs where the outage probability of different hops are comparable, the PDF of the FSO-based hops affects the network performance.

\emph{On the effect of RF-based transmit antennas:} Figure 4 demonstrates the effect of the number of RF transmit  antennas on the network outage probability. Also, the figure compares the system performance in the cases with short and long codewords, i.e., in the cases with small and large values of $C_i,\tilde C_i.$ Here, we consider Gamma-Gamma distribution of the FSO-based hops, ideal PAs, $R_i=1.5$ npcu, $M_i=1, T_i=\tilde T_i=1, \forall i,$ and $\text{SNR}=10$ dB. As seen, with short codewords, the outage probability decreases with the number of RF-based transmit antennas monotonically. With long codewords, however, the outage probability is (almost) insensitive to the number of antennas for $N_i\ge 3$. This is because, with the parameter settings of the figure, the data is correctly decoded with high probability as the number of antennas increases. Finally, the outage probability decreases with $C_i,\tilde C_i,$ because the HARQ exploits time diversity as more channel realizations are experienced within each codeword transmission.

\vspace{-2mm}
\section{Conclusion}
We studied the performance of RF-FSO multi-hop networks using long codewords. Considering different channel conditions, we derived closed-form expressions for the network outage probability in the cases with and without HARQ. As demonstrated, there are mappings between the performance of RF-FSO based multi-hop networks and the ones using only the RF- or the FSO-based communication. Moreover, the HARQ can effectively improve the energy efficiency and compensate the effect of hardware impairments. Finally, while the outage probability decreases with the number of RF transmit antennas,  the network outage probability  is not sensitive to the number of RF antennas for long codewords. Extension of our results can be found in \cite{RFFSOmultitrans} where we analyze the network ergodic rate and the required number of antennas in different conditions.
\vspace{-2mm}
%

\begin{figure}
\centering
  \includegraphics[width=0.8\columnwidth]{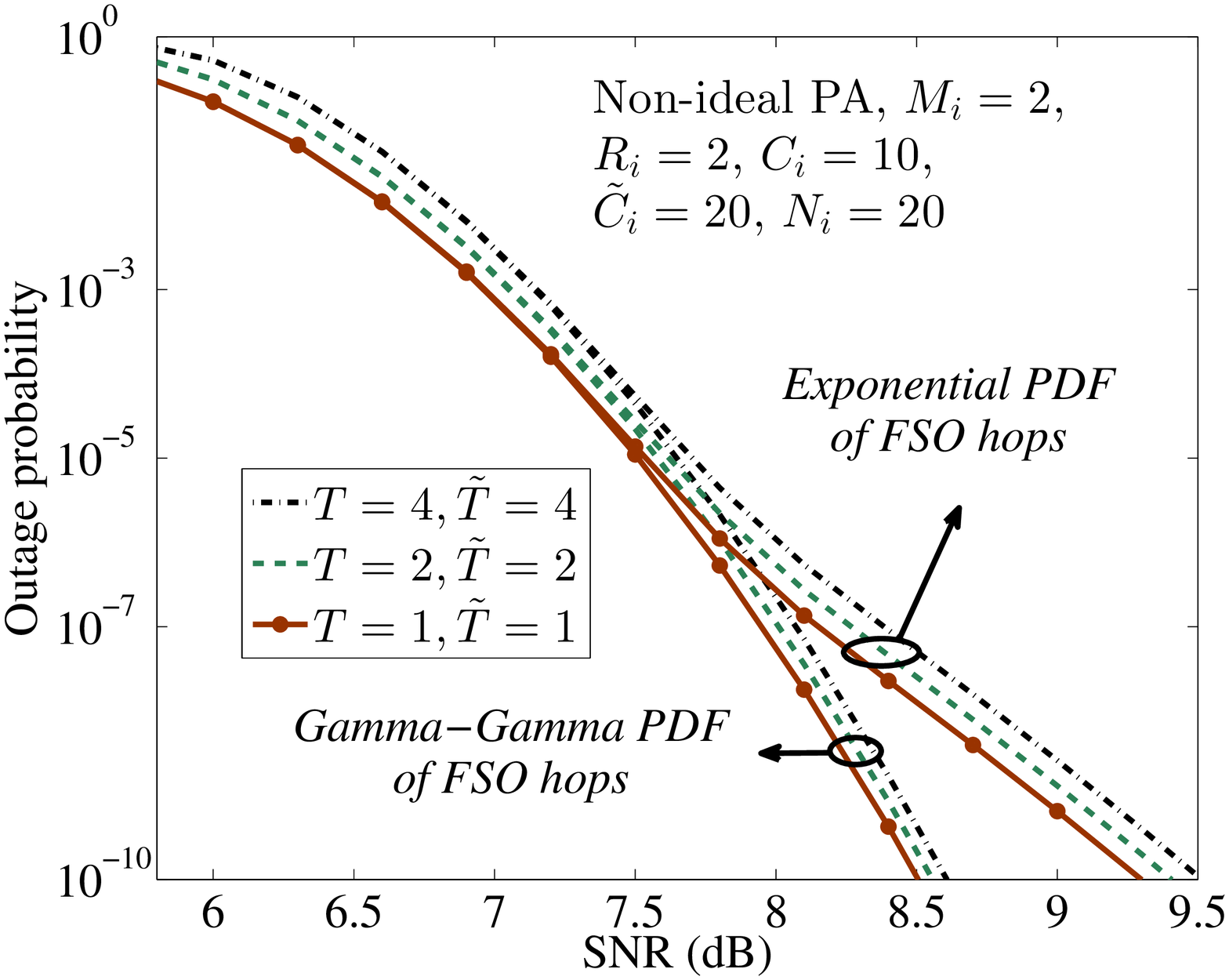}\\\vspace{-3mm}
\caption{The outage probability for different numbers of RF- and FSO-based hops, i.e., $T$ and $\tilde T$. Non-ideal PA, $\vartheta_i=0.5, \epsilon_i=0.75, P_i^\text{max}=25$ dB, $M_i=2, N_i=20, R_i=2$ npcu, $C_i=10,$ and  $\tilde C_i=20, \forall i.$}\vspace{-4mm}\label{figure111}
\end{figure}

\begin{figure}
\centering
  \includegraphics[width=0.8\columnwidth]{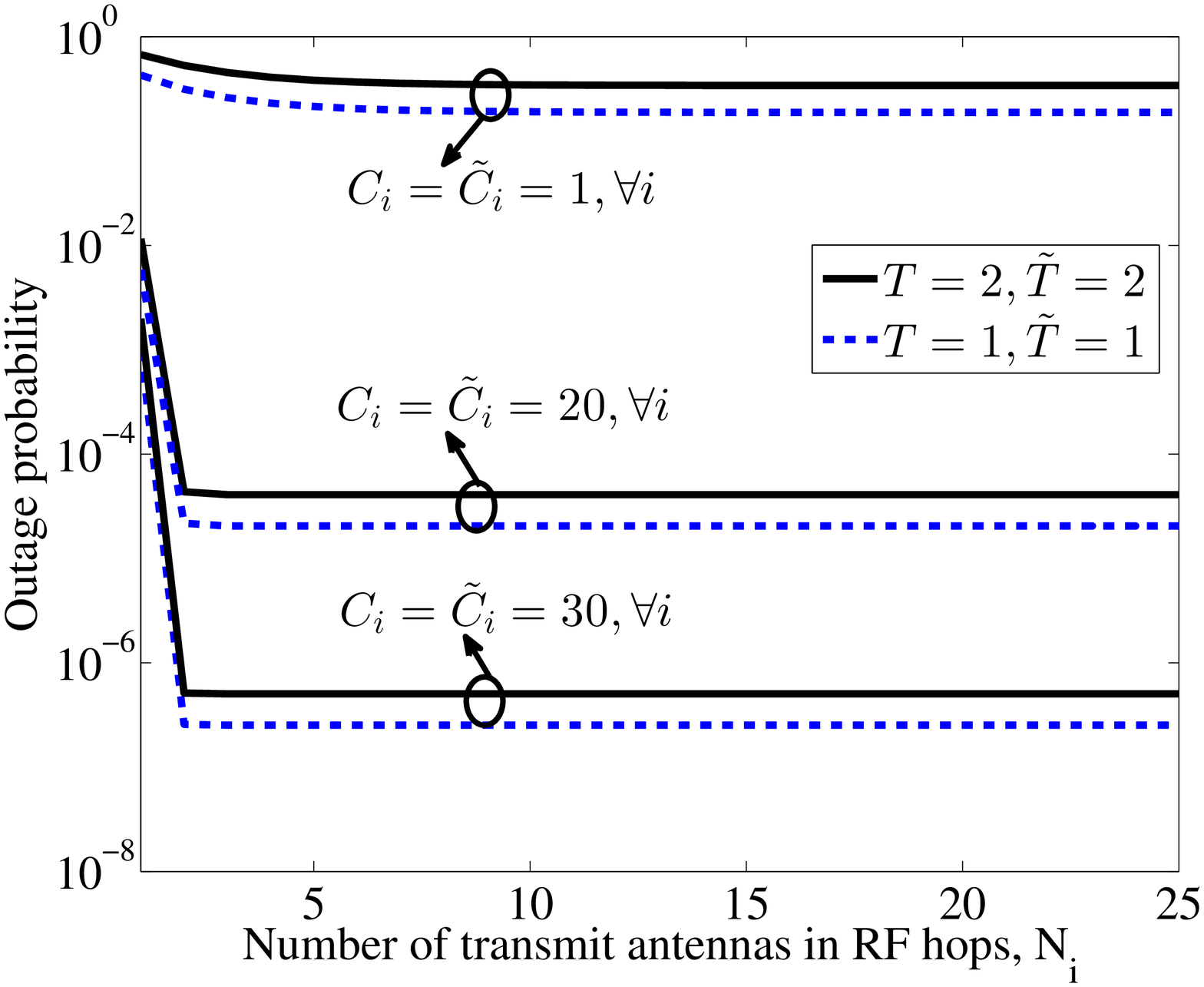}\\\vspace{-3mm}
\caption{Outage probability for different number of transmit antennas in the RF-based hops. Gamma-Gamma distribution of the FSO-based hops, ideal PA, $R_i=1.5$ npcu, $M_i=1, T_i=\tilde T_i=1, \forall i,$ and $\text{SNR}=10$ dB. }\vspace{-5mm}\label{figure111}
\end{figure}

%

\vspace{-0mm}
\section*{Acknowledgement}
The research leading to these results received funding from the European Commission H2020 programme under grant agreement $n^{\circ}$671650 (5G PPP mmMAGIC project), and from the Swedish Governmental Agency for Innovation Systems (VINNOVA) within the VINN Excellence Center Chase.
\bibliographystyle{IEEEtran} 
\bibliography{masterFSO3}
\vfill
\end{document}